\documentclass[aps,pre,twocolumn,showpacs]{revtex4-2}
\usepackage{graphicx}  
\usepackage{dcolumn}   
\usepackage{bm}        
\usepackage{amssymb,amsmath}   
\usepackage{float}
\usepackage[section]{placeins}
\usepackage{hyperref}
\usepackage{comment}
\usepackage{enumitem}

\newcommand{\Tt}{\langle 2,2 \rangle}

\newcommand{\Wnxnz}{\mathcal{W}_{\bar{x}\bar{z}}}
\newcommand{\Pnxnz}{\mathcal{P}_{\bar{x}\bar{z}}}
\newcommand{\Pxnz}{\mathcal{P}_{x\bar{z}}}
\newcommand{\Wnz}{\mathcal{W}_{\bar{z}}}
\newcommand{\Wxnz}{\mathcal{W}_{x\bar{z}}}

\usepackage[ruled]{algorithm}
\usepackage{algpseudocode}

\usepackage{xcolor}
\definecolor{mygray}{gray}{0.6}

\usepackage[most]{tcolorbox}  

\begin{document}

\title{High Resolution Study of the 2D ANNNI Model Using a Two-replica Cluster Algorithm and Population Annealing}
\author{Shane Keiser}
\affiliation{Department of Physics, University of Massachusetts, Amherst, Massachusetts 01003 USA}
\affiliation{Department of Nuclear Engineering and Radiological Sciences, University of Michigan, Ann Arbor, Michigan 48109 USA}
\author{Jonathan Machta}
\affiliation{Department of Physics, University of Massachusetts, Amherst, Massachusetts 01003 USA}
\affiliation{Santa Fe Institute, 1399 Hyde Park Road, Santa Fe, New Mexico 87501, USA}

\graphicspath{{figures}}

\date{\today} 

\begin{abstract}
The axial next-nearest-neighbor Ising (ANNNI) model in two dimensions is studied using population annealing combined with a two-replica cluster algorithm.  We are able to fully resolve the sequence of sharp specific heat peaks that characterize the finite-size incommensurate floating phase.  We also show that the two-replica cluster algorithm is much more effective in equilibrating the system than either single-replica cluster methods or the Metropolis algorithm when these are combined with population annealing.  We argue that effectiveness of the new algorithm is due to its ability to move groups of defect lines between replicas combined with resampling in population annealing, which removes replicas from the population that have larger numbers of defect lines.
\end{abstract}

\maketitle

\section{Introduction}

Materials with short range attractive and long range repulsive potentials often display equilibrium order with discrete translational symmetries.  If the interactions are anisotropic, layered structures may form.  Perhaps the simplest example of layering is found in the anisotropic next nearest neighbor Ising (ANNNI) model \cite{fisher_selke_1980,selke_annni_1988}.  The ANNNI model has both ferromagnetic  nearest neighbor interactions and competing antiferromagnetic next-nearest-neighbor (NNN) interactions in the axial  direction. The ANNNI model has a rich phase diagram that features layered structures at low temperatures.  

We  studied the 2D ANNNI model in the region in parameter space where the NNN interactions are sufficiently strong that the low temperature phase  displays layered ordering with two spin-up layers followed by two spin-down layers in the axial direction.  At higher temperatures there is believed to be an {\it incommensurate floating} (IC) phase with quasi-long-range order that is layered but with defect lines.  Finally, at high temperature, there is a paramagnetic phase with short-range order. Although there are theoretical reasons \cite{villain_bak_1981} for expecting an IC phase, there is no proof of its existence, and the numerical results have been inconclusive, with disagreement on the existence \cite{derian_modulation_2006, chandra_floating_2007} of the IC phase and its width in temperature \cite{hu_resolving_2021}.  In this work, we find strong evidence for the existence of an IC phase but are unable to locate the transition temperature due to strong finite size corrections.

The ANNNI model is difficult to simulate because of the frustration induced by the competing ferromagnetic and antiferromagnetic interactions. In addition to the standard Metropolis algorithm, several other computational methods have been applied to study the ANNNI model. In 3D, near the uniaxial Lifshitz point, Henkel and Pleimling applied a generalization of the Wolff algorithm \cite{wolff_collective_1989} for interactions of both signs to the ANNNI model \cite{PhysRevLett.87.125702}. However, the Henkel--Pleimling cluster algorithm is not useful in the regime of strong NNN interactions for reasons discussed in \cite{zheng_communication_2022} and in the present paper. The most successful computational methods to date for the ANNNI model use transfer matrices, either alone \cite{derian_modulation_2006,hu_resolving_2021} or in combination with Monte Carlo moves \cite{matsubara_sato_koseki_1997,sato_matsubara_1999,matsubara_domain_2017}. In the present work, we introduce a new computational method that uses population annealing \cite{hukushima_population_2003,machta_population_2010,WaMaKa15b} together with two-replica cluster moves~\cite{swendsen_replica_1986,ChMaRe98b,redner_TR_1998,Li_two_replica2001} and single-spin flips. We find that this method is far more effective at equilibrating the 2D ANNNI model than either the Metropolis algorithm or the Wolff--Henkel--Pleimling algorithm in the strongly frustrated regime. 


In Sec.~\ref{sec:model} we introduce and discuss the relevant features of the ANNNI model. Section \ref{sec:methods} reviews population annealing and the two-replica cluster method, and describes how they are combined. This section also reviews the competing algorithms and introduces the observables to be measured. Results for both ANNNI model physics and algorithmic performance are presented in Sec.\ref{sec:results}. The paper closes with a discussion, Sec.~\ref{sec:discuss}.

\section{Model}
\label{sec:model}

The ANNNI model energy, $E$ is described by
\begin{align}
    E = -\left(\sum_{\langle i,j \rangle} \sigma_i \sigma_j  - \kappa \sum_{ [i,j]_\text{z}} \sigma_i \sigma_j \right),
\end{align}
\noindent where the sums are over nearest neighbors, $\langle i,j \rangle$ and next-nearest neighbors in the axial $z$-direction, $[i,j]_\text{z}$. Our simulations are carried out on $L \times L$ square lattices with periodic boundary conditions. The system size, $L$ is chosen to be a multiple of 4 so that it admits layered structures without defects.  

 At zero temperature and in any dimension $d$, as $\kappa$ increases from zero, there is a transition from the twofold degenerate ferromagnetic ground state with energy per spin of $(-d+\kappa)$ to a fourfold degenerate, $\Tt$ ground state. The $\langle 2,2\rangle$ ground state displays a repeating pattern $\uparrow \uparrow \downarrow \downarrow$ in the axial direction. The four degenerate $\langle 2,2\rangle$ ground states differ by translations in the axial direction and have the energy per spin $(-d+1-\kappa)$. Comparing these energies, one sees that there is a transition from the ferromagnetic ground state to the $\langle 2,2\rangle$ ground state at $\kappa=1/2$.

Although the ground state is the same in any number of dimensions, the 2D and 3D phase diagrams are quite different. In 3D there are many ordered phases characterized by different repeating patterns in the axial direction. In 2D, the subject of this study, it is believed that there are four distinct phases. For sufficiently low temperature and $\kappa<1/2$ there is a ferromagnetic phase, while for $\kappa>1/2$ there is an ordered $\Tt$ phase. At high temperature there is a paramagnetic phase. Intervening between the $\Tt$ phase and the paramagnetic phase there is believed to be a narrow critical phase, known as the {\it incommensurate floating} (IC) phase, which displays quasi-long range order. The transition between paramagnetic and IC phases is believed to be in the Brezinski--Kosterlitz--Thouless universality class, though the location of the transition line is poorly determined. The transition between the IC phase and the ordered phase is believed to be of the Pokrovsky--Talapov (PT) \cite{pokrovsky_talapov_1979, selke_annni_1988} incommensurate to commensurate type and is described by free fermion theory. 

\begin{figure}[h!]
    \centering
    \includegraphics[width=\columnwidth]{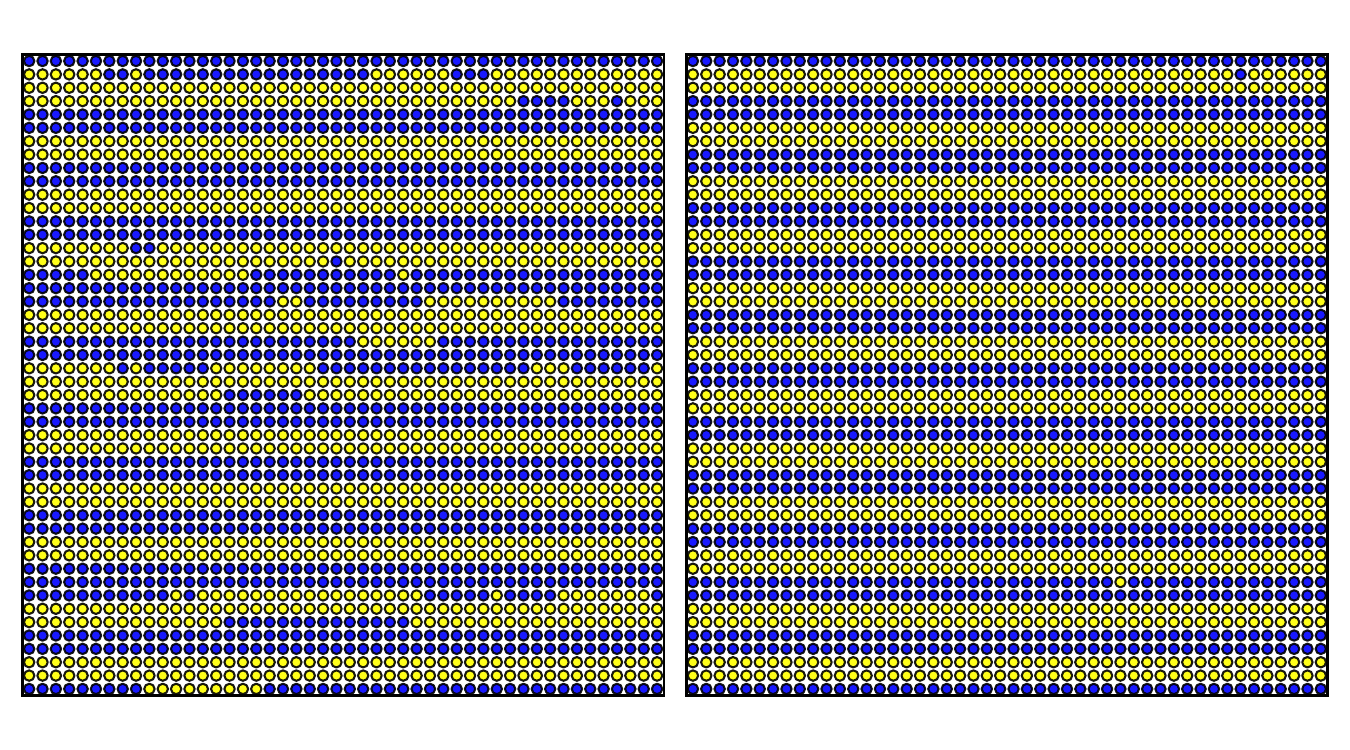}
    \caption{Snapshots of two $L=48$ replicas at $\beta=1.09$  near the last specific heat peak with coexistence between a configuration in the IC phase  (left panel) with four defect lines ($k = 11/48$) and a configuration in the $\Tt$ phase ($k = 12/48$).}
    \label{fig:snapshots}
\end{figure}

In the IC phase, typical configurations have regions of $\Tt$ order separated by defect lines consisting of  layers with three or more like spins. A three-spin defect line translates the $\Tt$ phase type by one. After four such defect lines, the $\Tt$ phase returns to the same translation as before. Thus, in periodic boundary conditions with $L$ a multiple of four, as is the case in our study, the number of three-spin defect lines must be a multiple of four. Two typical snapshots taken from our simulations are shown in Fig.~\ref{fig:snapshots}. These snapshots are taken near the transition between the IC and $\Tt$ phases. The left panel of Fig.\ \ref{fig:snapshots} shows a configuration in the IC phase with four defect lines. The right panel shows a configuration that is representative of the $\Tt$ phase and features only point defects. The fact that widely separated defect lines come and go in groups of four makes it very difficult to equilibrate the ANNNI model in the IC phase \cite{matsubara_domain_2017}.

\section{Computational Methods}
\label{sec:methods}

All simulation results reported in this work are carried out within the population annealing framework. Population annealing (PA) is a sequential Monte Carlo method that is closely related to simulated annealing. Like simulated annealing, PA uses as a subroutine an equilibrating Monte Carlo method, such as the Metropolis algorithm. The basic idea for both of these annealing frameworks is to improve the performance of the Monte Carlo method at low temperature by first annealing from high temperature. Population annealing is discussed in Sec.~\ref{sec:PA}.

In this work we use three equilibrating Monte Carlo methods, the Metropolis single spin flip algorithm, the Wolff cluster algorithm as adapted by Pleimling and Henkel for the ANNNI model \cite{PhysRevLett.87.125702}, and a two-replica cluster algorithm based on Refs.~\cite{ChMaRe98b, redner_TR_1998, Li_two_replica2001}.  
Details for the two-replica cluster algorithm are given in Sec.~\ref{sec:TR}.

\subsection{Population Annealing Monte Carlo}
\label{sec:PA}
Population Annealing \cite{machta_population_2010,amey_analysis_2018} is a sequential Monte Carlo method that is useful for simulating systems with frustration and rough free energy landscapes, such as the ANNNI model. In the framework of PA, a population of $R$ replicas is initialized and equilibrated at high temperature (small $\beta$). Subsequently, $\beta$ is increased according to an annealing schedule. Like simulated annealing, at each step in the schedule, each replica is acted on by a Markov Chain Monte Carlo (MCMC) algorithm that converges to the equilibrium distribution at the current $\beta$. In population annealing, after the action of the MCMC, the population undergoes resampling (differential reproduction), in a way that brings the population closer to the equilibrium distribution at the new inverse temperature $\beta^\prime$. 
This is accomplished by first computing the reweighting factor $\tau_r(\beta,\beta^\prime)$ for replica $r$,
\begin{equation}
\tau_r(\beta,\beta^\prime) = \frac{R}{R_\beta}\frac{e^{-(\beta'-\beta)E_r}}{Q(\beta,\beta')},
\end{equation}
where $E_r$ is the energy of replica $r$ and $R_\beta$ is the population size at inverse temperature $\beta$. The normalization factor $Q$ and the population size ratio, $R/R_\beta$ are required to maintain the population size near the target value of $R$.  The normalization factor is given by
\begin{equation}
\label{eq:Q}
Q(\beta,\beta') = \frac{1}{R_\beta}\sum_{r=1}^{R_\beta}e^{-(\beta-\beta')E_r},
\end{equation}
and provides an estimator of free energy differences \cite{machta_population_2010},
\begin{equation}
\label{eq:bF}
    \beta^\prime F(\beta^\prime) - \beta F(\beta) = - \log Q(\beta,\beta').
\end{equation}
 
Resampling is implemented by choosing a random whole number $n_r$ from a distribution whose mean is $\tau_r(\beta,\beta^\prime)$. The initial population at inverse temperature $\beta^\prime$ then has $n_r$ copies of replica $r$. If $n_r=0$, replica $r$ is culled from the population, while if $n_r>1$ then replica $r$ is reproduced. There are many ways to choose $n_r$ \cite{gessert_resampling_2023}. Here we use nearest integer resampling, $n_r(\beta,\beta')= \lfloor \tau_i(\beta,\beta') \rfloor$ with probability $\lceil \tau_i(\beta,\beta') \rceil - \tau_i(\beta,\beta')$ and $n_i(\beta,\beta')= \lceil \tau_i(\beta,\beta') \rceil$ otherwise. This method minimizes the variance in $n_r$ at the expense of allowing for order $\sqrt{R}$ fluctuations in the population size.

A distinctive feature of sequential Monte Carlo methods as opposed to MCMC methods is that the average values of observables are obtained from population averages rather than time averages so that the estimator $\tilde{A}_\beta$ of observable $A$ at inverse temperature $\beta$ is the population mean, $\tilde{A}_\beta=(1/R_\beta)\sum_{r} A_r$. 

The value of the dimensionless free energy estimator $\beta F(\beta)$ can be computed using thermodynamic integration, by summing Eq.\ \eqref{eq:bF}. The free energy is useful for several purposes. The first important use is to compute weighted averages of observables obtained from multiple runs. If one carries out $M$ PA runs and computes both $\beta F(\beta)$ and the estimator of an observable $\tilde{A}_m$ for run $m$ then the best estimator $\bar{A}$ from these runs is not the ordinary average, but the weighted average \cite{machta_population_2010},
\begin{equation}
    \bar{A} = \frac{\sum_{m=1}^M \tilde{A}_m R_m e^{-\beta F_m}}{\sum_{m=1}^M  R_m e^{-\beta F_m}}.
\end{equation}
The weighting factor is the product of the population size of the run $R_m$ and the exponential of the absolute value of the free energy $\beta F(\beta)_m$. For the case of the free energy itself, the best estimator from several runs is 
\begin{equation}
\label{eq:barF}
    \beta \bar{F} =-\log \left[ \sum_{m=1}^M  e^{-\beta F_m}\right]
\end{equation}
The free energy is also useful for comparing the performance of different MCMC algorithms used in PA  since the algorithm with the lower free energy is closer to equilibrium. The variance of the dimensionless free energy provides a useful internal indicator of equilibration \cite{WaMaKa15b}. Systematic errors in estimating the free energy are proportional to the run-to-run variance of the free energy estimator.
However, this variance measure may be misleadingly small if important regions of phase space have not been sampled.

\subsection{Two--Replica Cluster Algorithm}
\label{sec:TR}

Two-replica cluster methods were first introduced by Swendsen and Wang \cite{swendsen_replica_1986} and developed by several authors \cite{redner_TR_1998,  houdayer_cluster_2001,jorg06,MaStNe07}. Here we use the method developed in Ref.~\cite{redner_TR_1998}. Two-replica methods have primarily been applied to spin glass system with mixed success \cite{swendsen_replica_1986,houdayer_cluster_2001,jorg06,MaStNe07,Zhu_cluster_2015,munster_weigel_2026,chilin2026clustermovesentropicreservoir} . The method was shown to be very efficient for the simpler case of an Ising antiferromagnet in an external field \cite{redner_TR_1998, Li_two_replica2001} where single-replica cluster techniques fail. This success motivated us to apply the method to the ANNNI model.

We first discuss single-replica cluster algorithms such as the Swendsen-Wang or Wolff algorithms before turning to closely related two-replica methods. In single-replica cluster algorithms, as applied to the Ising model, cluster(s) of spins are stochastically constructed and then the whole cluster is flipped with some probability. A bond connecting two spins in the same cluster must be `satisfied' in the sense that the interaction energy is increased when one of the two spins is flipped. Bonds connecting two spins in the same cluster are referred to as `occupied.' When the cluster is flipped, all occupied bonds remain satisfied but the energy of the system changes due to bonds on the perimeter of the cluster since satisfied bonds become unsatisfied and vice versa. The stochastic rule for building a cluster by occupying satisfied bonds must obey detailed balance with respect to the energy change on the perimeter of the cluster when it is flipped. The detailed balance condition for single replica cluster methods applied to Ising models yields the bond occupation probability $p^{(1)}_{ij}$,
\begin{equation}
    p^{(1)}_{ij} = 1 - \exp(-2\beta J_{ij}).
\end{equation}
In the case of the ANNNI model, $J_{ij}$ is either $1$ for nearest-neighbor spins or $-\kappa$ for next-nearest-neighbor spins in the $z$ direction. In the Swendsen-Wang algorithm, all satisfied bonds are occupied with the above probability, and then each cluster is flipped with probability $1/2$. In the Wolff algorithm, a single cluster is grown from a randomly chosen seed site and flipped with probability one.

In the two-replica methods, cluster(s) of {\it pairs} of spins are constructed stochastically from two independent copies of the system. The two spins in the pair reside at the same site in different copies of the system. In the construction of cluster(s), occupied bonds in the cluster must be satisfied in both replicas. When a cluster is flipped, spins in both replicas are flipped, so the detailed balance condition requires the occupation probability,
\begin{equation}
    p^{(2)}_{ij} = 1 - \exp(-4\beta J_{ij}).
\end{equation}
Note the factor of two difference between the single- and two-replica occupation probabilities. As is the case for single-replica cluster methods we can either build all clusters or a single cluster from a random seed site. In the present work we follow the Wolff single-cluster paradigm. 

A single MC sweep for both the Wolff algorithm and the two-replica cluster algorithm is defined as the number of cluster flips needed to change every spin on average once. Since our annealing schedules use small temperature steps, statistics from the previous annealing step determine how many clusters to build in the current step.

There is an important conceptual difference between single-replica  and two-replica cluster methods. This difference is best understood in the versions of these algorithms in which all clusters are built. Consider the case of the single-replica (Swendsen--Wang) algorithm with ferromagnetic couplings ($J_{ij}>0$). Then all spins in the cluster have the same sign while spins in different clusters are independent. Thus the spin-spin correlation function is equal to the probability that the two spins are in the same cluster, and the onset of a 'giant' cluster with non-vanishing density is equivalent to the onset of spontaneous magnetization. The dramatic reduction in critical slowing achieved by the Wolff and Swendsen--Wang algorithms is closely associated with the existence of clusters of all sizes in the critical regime. Unfortunately, all of these advantageous features of the single-replica cluster algorithms break down for systems with antiferromagnetic bonds, since a satisfied antiferromagnetic bond will have opposite spins at its two ends and spins in the same cluster may not have the same sign. Single-replica cluster algorithms applied to frustrated systems generally have the onset of a giant cluster at a temperature above the ordering temperature and, therefore, perform poorly in the critical and low temperature phases.

Now consider two-replica cluster algorithms. The site overlap, $q_i$, at site $i$ is defined by $q_i=\sigma_i\tau_i$ where $\sigma$ and $\tau$ refer to the spins in two independent replicas. Since all bonds in a two-replica cluster are satisfied in both replicas, all sites in a cluster have the same value of the overlap. It is important to note that the constancy of the overlap in a two-replica cluster is independent of whether there are antiferromagnetic occupied bonds in the cluster. The overlap is known to be a useful order parameter for frustrated systems as first noted in the case of spin glasses by Edwards and Anderson \cite{Edwards_1975}. Unfortunately, it is not true for the two-replica cluster algorithm that $\langle q_iq_j\rangle$ is given by the probability of being in the same cluster or that the onset of a giant cluster necessarily coincides with a non-zero Edwards--Anderson order parameter \cite{redner_TR_1998}. In particular, it is possible that there are two giant clusters with equal density and opposite overlap so that the average overlap is zero and that the onset of order occurs when the density of the two giant clusters differs \cite{MaStNe07}. This scenario is the case for spin glasses with the result that two-replica algorithms are not dramatically more efficient than single-spin flip algorithms. As we shall see below, two-replica cluster methods do have substantial advantages for the ANNNI model.

Finally, we note that two-replica cluster flips are well-suited for combining with PA because on each sweep a different pairing of replicas may be used. However, it is important in this pairing that the replicas be nearly independent. This is achieved using a method introduced in Ref.~\cite{weigel_understanding_2021}, section VB. In this method, replicas that are copies of one another from resampling are placed next to one another in the replica list. To prevent closely correlated replicas from being paired in a two-replica move, we forbid pairing for replicas closer than 10 neighbors in the replica list.

\subsection{Simulation Details}

A single MC step of the two-replica cluster algorithm for replicas $\boldsymbol{\sigma}$ and $\boldsymbol{\tau}$ is composed of the following instructions:
\begin{enumerate}
    \item Select a root site at random to seed the cluster. The spins at the site being checked in $\boldsymbol{\sigma}$ and $\boldsymbol{\tau}$ are denoted respectively as $\sigma$ and $\tau$.
    \item Check (serially) all neighboring spins of $\sigma$ and $\tau$, denoted $\sigma^\prime$ and $\tau^\prime$, in a breadth--first search, and add them to the cluster if the bond criterion is satisfied in both replicas:
    \begin{enumerate}
        \item For nearest-neighbors, if $\sigma \sigma^\prime = \tau\tau^\prime = 1$, add to the cluster with probability $p = 1 - \exp(-4\beta)$
        \item For next-nearest-neighbors in the axial direction, if $\sigma \sigma^\prime = \tau\tau^\prime = -1$, add to the cluster with probability $p = 1 - \exp(-4\beta\kappa)$
    \end{enumerate}
    \item Repeat step 2 with the site that was most recently added to the cluster and is not yet checked
    \item When all sites in the cluster have been checked and no new sites are added, flip the entire cluster in both replicas.
\end{enumerate}

To comprehensively check the neighbors of all relevant sites in a breadth-first-search, we append the site's coordinates to a stack when it is added to the cluster. We also flip a `checked' boolean to true. Thereafter, the top site is pulled from the stack, and its neighbors are checked. The cluster stops growing when the stack is empty \cite{newman_monte_2010}.

For our main simulations, we use PA with a combined with two-replica cluster and Metropolis sweeps, which we refer to as the TR algorithm. Experiments showed that while using only two-replica moves is satisfactory, performance is improved by the addition of a small number of Metropolis sweeps at each temperature step. The pseudo-code for the TR algorithm is shown below:

\begin{algorithm}[H]
\renewcommand{\thealgorithm}{}
    \begin{algorithmic}[1]
        \State Define annealing schedule $(\beta_0=0, \cdots \beta_N)$
        \State Initialize $R_0$ replicas with side length $L$ at $\beta_0$
        \For{$i \gets 1$ to $N$} \Comment \textcolor{mygray}{Annealing loop}
            \State Resample population and increase inverse 
            \Statex \hspace{12pt} temperature ($\beta_{i-1} \rightarrow \beta_i$)
            \For{$j \gets 0$ to $S$}
                \State Pair replicas randomly
                \For{$k \gets 0$ to $R/2$}
                    \State Execute 1 sweep of Two--Replica 
                    \Statex \hspace{39pt} cluster algorithm for replica pair $k$
                \EndFor
                \EndFor
                \For{$k \gets 0$ to $R$}
                    \State Execute $\lfloor{S/10}\rfloor$ sweeps of Metropolis algorithm
                \EndFor
                \State Collect data
        \EndFor
    \end{algorithmic}
    \caption{Two--Replica Cluster Algorithm with Population Annealing}
    \label{alg:tr-pa}
\end{algorithm}

At each temperature we do $S$ two-replica cluster sweeps followed by $S/10$ Metropolis sweeps. The simulation begins at infinite temperature, $\beta = 0$, and while more optimized annealing schedules exist \cite{amey_analysis_2018,Barzegar_opt_schedule2024}, we use a simple ad-hoc annealing schedule
\begin{align*}
    \Delta \beta = \begin{cases}
        0.005, &\, \beta < 0.6\\
        0.0005, &\, 0.6 \leq \beta < 0.93\\
        0.0001, &\, 0.93 \leq \beta \leq 1.2,
    \end{cases}
\end{align*}
\noindent which increases the resolution in the temperature region of the finite-size IC phase where there are rapidly varying observables, such as multiple sharp peaks in the specific heat. This schedule includes 3477 temperature steps. For $L=256$, we found that the system is poorly equilibrated for $\beta>0.9$ so we do not use that data in Sec.~\ref{sec:algoresults}.

Unless otherwise stated, we use the hyperparameters reported in Table \ref{tab:hyperparams} and all results are for $\kappa=0.6$ in the regime where the low temperature phase has $\Tt$ ordering.

\begin{table}[htb]
\caption{Simulation hyperparameters.}
\label{tab:hyperparams}
\centering
\begin{ruledtabular}
\begin{tabular}{lccccc}
Size, $L$ & 48 & 64 & 96 & 128 & 256\\
Sweep count, $S$ & 100 & 100 & 100 & 200 & 50\\
Target population size, $R$ & 6400 & 6400 & 6400 & 3200 & 3200 \\
Independent run count, $M$ & 100 & 100 & 100 & 100 & 60 \\
\end{tabular}
\end{ruledtabular}
\end{table}
    
\section{Observables}
We measure various quantities, some to elucidate the physics of the ANNNI model and others to understand the behavior of the algorithms. All quantities are averaged over replicas (or pairs of replicas) in each run, and then weighted averages are obtained from the full set of runs. Error bars are obtained by bootstrapping the weighted averages. 

The specific heat, $c$ is obtained from the variance of the energy. Care must be taken in performing weighted averages \cite{ebert_weighted_ave_2022}. Weighted averaging is first done separately for the first and second moments of the energy and then combined to obtain the variance,
\begin{equation}
\label{eq:C}
    c= \frac{\beta^2}{N} \left[\overline{E^2}- (\bar{E})^2\right].
\end{equation}

In the finite-size IC phase, typical configurations consist of regions of the four $\Tt$ phases separated by defect lines in the transverse direction. The structure of these states can be quantified by measuring the line magnetization, $m_{\ell}$,
\begin{equation}
    m_{\ell}(z)=\frac{1}{L}\sum_{x=1}^L s_{(x,z)},
\end{equation}
in each replica where $z$ is the axial coordinate and $x$ the transverse coordinate. We perform an FFT on $m_{\ell}(z)$ for each replica and identify the dominant wavevector $k$, defined as the wavevector with the largest weight. Here we define wavevectors as the inverse wavelength, $k=n/L$ where $n$ is a whole number and $L$ is the system size, {\it i.e.}\ we omit the factor of $2\pi$. For example, in the $\Tt$ phase $k=1/4$ since the repeat pattern, $++--$ has wavelength four.  

At each temperature step during data collection, we determine each replica's dominant wavenumber. From this we record the fraction $P_k$ of the population that occupies each wavenumber, as well as each wavenumber's average energy, $E_k$ and specific heat contribution, $c_k$.
The $c_k$ are defined as in Eq.~\ref{eq:C} with a factor of $P_k$ and the law of variances yields an expression for the components comprising the total specific heat, 
\begin{equation}
\label{eq:Cvk}
    c = \frac{\beta^2}{N}\left[\left(\sum_k c_k\right)  + \mathrm{Var}_{P_k}\left[ E_k\right]\right].
\end{equation}

We also calculate statistics of overlap $q= \sum_i q_i$ and the average is taken over pairs of replicas. (\cite{shirakura_kosterlitz-thouless_2014} From the overlap distribution we compute the overlap Binder cumulant,
\begin{equation}
    \label{eq:binder}
    B = \frac{1}{2}\left[3-\frac{\langle q^4\rangle}{ \langle q^2\rangle^2 }\right].
\end{equation}

Finally, using Eqs.~\eqref{eq:Q}, \eqref{eq:bF}, and \eqref{eq:barF}, we measure the free energy estimator and its run to run variance as a function of $\beta$. The primary purpose of these quantities is to evaluate equilibration and compare the performance of different algorithms and hyperparameter settings.

In order to better understand the behavior of the Wolff and TR algorithms we consider the probability and average size of clusters conditioned on whether they wrap in the $x$ or $z$ directions. For example, the quantity $\Pxnz$ is the probability that a cluster wraps in the $x$ direction but not in the $z$ direction, while $\Wxnz$ is the average size of such clusters.

\section{Results}
\label{sec:results}
In this section we discuss the results of the simulations and conclusions we draw from them. Section \ref{sec:ANNNIresults} gives results concerning the equilibrium behavior of the ANNNI model while Sec.~\ref{sec:algoresults} gives results pertaining to the performance of the computational methods and Sec.~\ref{sec:TRflips} provides an explanation for the efficacy of two-replica cluster flips.

\subsection{ANNNI Equilibrium Behavior}
\label{sec:ANNNIresults}
\begin{figure}[H]
    \centering
    \includegraphics[width=\columnwidth]{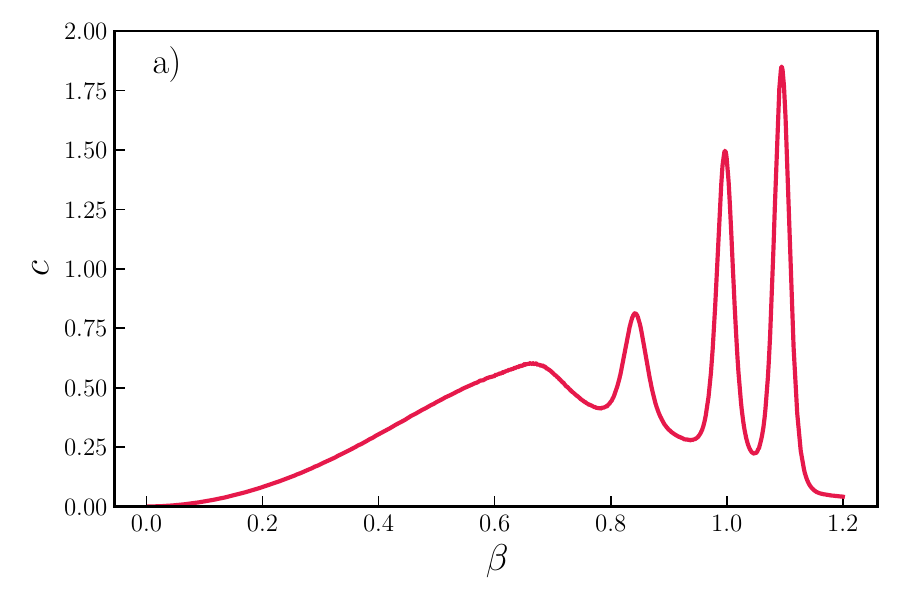}
    \includegraphics[width=\columnwidth]{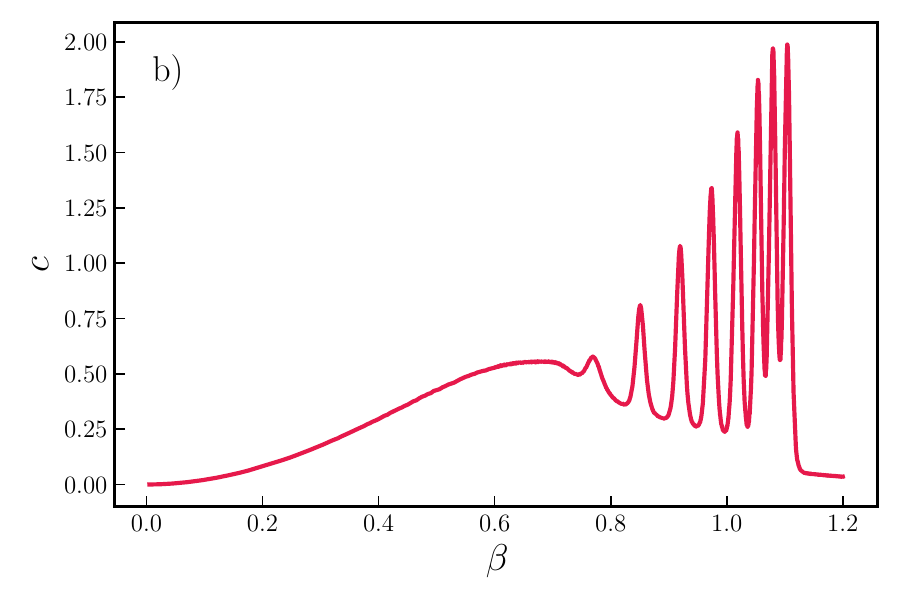}
    \caption{Specific heat, $c$, as a function of $\beta$, for sizes (a) $L = 48$ and (b) $L=128$.}
    \label{fig:C}
\end{figure}

To better understand the IC phase and the transition to the $\Tt$ phase we first consider the behavior of the specific heat.
Figure \ref{fig:C} shows the specific heat as a function of $\beta$ for system sizes $L=48$ and $128$. The notable feature of these plots is the sequence of sharp peaks with the number of peaks and their sharpness increasing with system size. These peaks correspond to first order-like transitions between states with different number of defect lines. Figure \ref{fig:snapshots} shows two snapshots at the last peak for $L=48$ and $\beta=1.09$ where there is coexistence between the IC and $\Tt$ phases. The left panel corresponds to a system with 4 three-spin defect lines ($k=11/48$)  and the right panel to a state with no defect lines ($k=12/48$) in the ordered phase. The region where the specific heat is enhanced by successive transitions between states with increasing wavevectors has some features like the IC phase in the thermodynamic limit, and we refer to as the finite-size IC phase or, when there is no confusion, as the IC phase.

We can analyze the heights of these peaks for the sizes studied here from the second term in Eq.\ \eqref{eq:Cvk} together with two simplifying assumptions: (1) the peak corresponds to the coexistence with equal probability of two states that differ by four three-spin defect lines, and (2) the energy of a defect line is proportional to the transverse length of the system $L_x$, $E_d \propto L_x$.  To compare with other results, we consider systems with arbitrary aspect ratio.
Since each transition is associated with the coexistence of states that differ by 4 defect lines, we see that the peak heights $C_{\mathrm peak}$ behave like  $C_{\mathrm peak} \propto  \beta L_x/L_z$ so, for a square system, the peak heights are independent of $L$ but more generally they are proportional to the aspect ratio.

Our numerical results for the specific heat can be compared with Refs.\ \cite{rastelli_specific_2010,hu_resolving_2021}.  Figure \ref{fig:C} is very similar to figure 2 of Rastelli et.~al.~\cite{rastelli_specific_2010} where systems of similar size in periodic boundary conditions are studied. The primary qualitative difference is that all peaks in our Fig. \ref{fig:C} are well-resolved, which is not the case in figure 2 of \cite{rastelli_specific_2010}.  
Hu and Charbonneau \cite{hu_resolving_2021} find much higher peaks that grow exponentially in $L_z$. This difference is likely because they use a transfer matrix approach, which creates a system with an infinite aspect ratio. Our simple approximation then predicts infinite peak heights.

\begin{figure}[h!]
    \centering
    \includegraphics[width=\columnwidth]{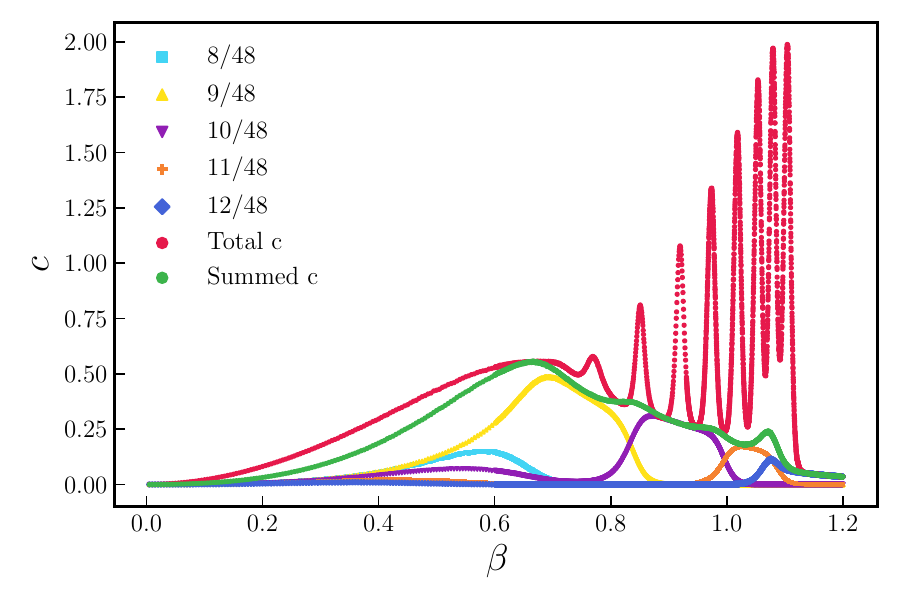}
    \caption{Contributions to the specific heat as a function of $\beta$ for system size $L=48$.  The total specific heat is the topmost (red) curve.  The sum of the specific heats from each wavevector state is second from the top (green). The contributions, $c_k$, to this sum successively appear and then disappear with increasing $\beta$. Contributions are shown only for wavevectors $k \geq 1/6$.}
    \label{fig:Cvk}
\end{figure}

In addition to contributions to the specific heat from the coexistence of defect states with different energies, there is a contribution, $c_k$ in Eq.\ \eqref{eq:Cvk}, from the energy fluctuations in a single state with wavevector $k$.   Figure \ref{fig:Cvk} shows the total specific heat, the specific heat contributions for each wavevector, $c_k$, and the sum of the contributions from each wavevector (the first term in Eq.\ \eqref{eq:Cvk}. We see that contributions from increasing wavevectors appear successively in $\beta$ and then drop out. The sum of all the contributions is relatively smooth and roughly follows the troughs of the total specific heat curve.

\begin{figure}[h!]
    \centering
    \includegraphics[width=\columnwidth]{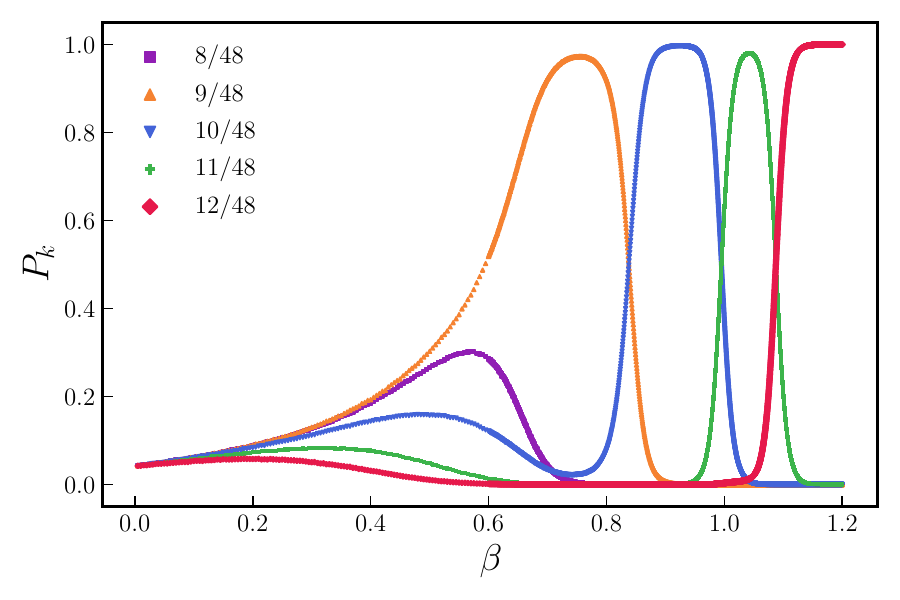}
    \includegraphics[width=\columnwidth]{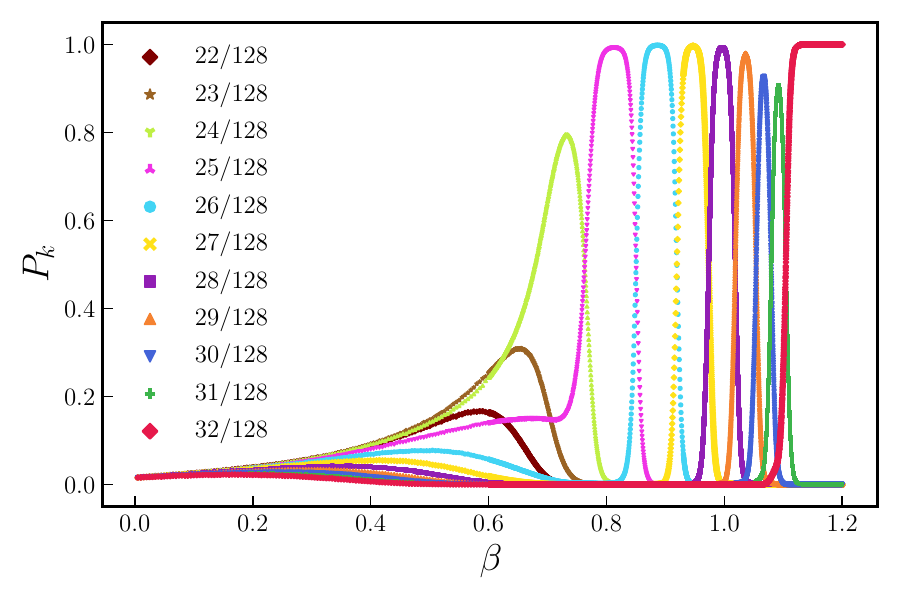}
    \caption{Probability $P_k$, of wavevector state $k$,  (a) $L = 48$ and (b) $L=128$ for wave vectors $k \geq 1/6$.}
    \label{fig:kfraction}
\end{figure}

Figure \ref{fig:kfraction} shows the estimated probability of being in a state with dominant wavevector $k$.  Corresponding to the specific heat results, successive wavevector states appear, reach a maximum, and then disappear as $\beta$ increases. Finally, for $\beta>1/T_{c2}$ the $k=1/4$ or $\Tt$ low temperature phase takes over. There is an interesting trend going from $L=48$ to $L=128$.  For $L=48$ the probabilities of the three largest wavevector states less than $k=1/4$ all reach probability nearly one.  On the other hand,  for $L=128$, the wavevector states immediately before $1/4$ do not reach one, implying that several wavevectors coexist throughout the region in $\beta$ near and below $\beta= 1/T_{c2}$.  This diversity of wavevector states was also observed in Ref.~\cite{matsubara_domain_2017}.  We believe that for large systems, the coexistence of multiple overlapping wavevector states persists throughout the IC phase and that, correspondingly, the specific heat curves will become smooth with troughs and peaks coming together, however, it is clear that even for $L=128$ the system is quite far from this regime.

\begin{figure}[h!]
    \centering
    \includegraphics[width=\columnwidth]{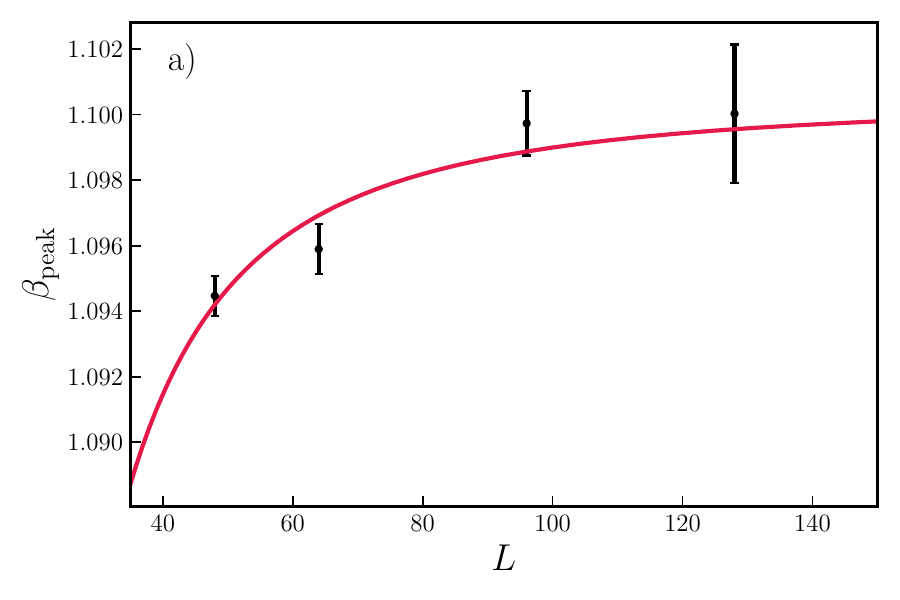}
    \hfill
    \includegraphics[width=\columnwidth]{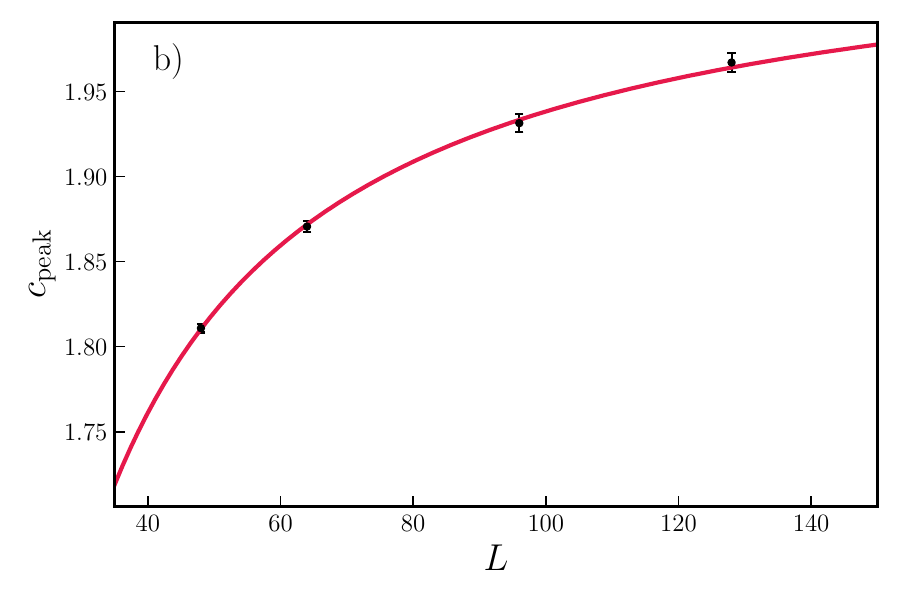}
    \caption{(a) Last specific heat peak position, $\beta_{\mathrm{peak}}$, as a function of system size, $L$.  The line is a fit for $\beta_{c2}$ of the form $\beta_{\mathrm{peak}}(L) = \beta_{c2} - a L^{-2}$ yielding $a = 15 \pm 4$ and $\beta_{c2}=1.101 \pm  0.001$  with $\chi^2/{\mathrm{df}} = 1.27$. (b) Last specific heat peak height, $c_{\mathrm{peak}}$ as a function of $L$. The line is a fit of the form $c_{\mathrm{peak}}=a/L - b$ yielding $a = -11.8 \pm 0.3$ and $b = 2.056 \pm 0.004$ with $\chi^2/{\mathrm{df}} = 0.52$.}
    \label{fig:extrapolateTc2}
\end{figure}

The location of the last peak is an estimator of the finite-size transition from the IC to the $\Tt$ phases. Figure \ref{fig:extrapolateTc2} shows the values of these peak locations as a function of system size together with an $L^{-2}$ extrapolation to the infinite size transition temperature, $T_{c2}$.  The result, $\beta_{c2}=1.101 \pm 0.001$ or $T_{c2}=0.908$, is in agreement with previous recent estimates \cite{matsubara_domain_2017,hu_resolving_2021}.  The error bars quoted above are statistical errors obtained from bootstrapping the weighted averages over runs. In Sec.~\ref{sec:algoresults}  we report on the equilibration of the TR algorithm and find that the $L=128$ runs are not  fully equilibrated and may have small systematic errors so the last phase transition occurs slightly late in $\beta$ which could lead to a slight underestimate of $T_{c2}$.
The $L^{-2}$ finite size correction in the fit is taken from free fermion theory and is seen to be a good fit with $\chi^2/{\mathrm{dof}}=1.53$. On the other hand, the exponent $2$ is poorly constrained by the data and $L^{-1}$ gives a slightly better fit,  $\chi^2/{\mathrm{dof}}=0.84$ with $T_{c2}=0.905$.  Free fermion scaling was also found in Ref.~\cite{Derian_2006}.

Our results for the specific heat near the transition display some but not other features of the PT transition. In agreement with the PT prediction, approaching the transition from the low temperature $\Tt$ phase there appears to be no singular behavior in $c$.  On the other, PT predicts a square root singularity approaching from the IC phase.  Figure \ref{fig:extrapolateTc2}b shows the peak height of the last peak as a function system size.  A fit to a finite peak height with a $1/L$ correction yields an asymptotic peak height $2.056 \pm 0.004$ with 
$\chi^2/{\mathrm{dof}}=0.52$.  Fits to square root and logarithmic singularities yields $\chi^2/{\mathrm{dof}}$ of 19 and 7, respectively. One explanation for the disagreement is that even for size 128 there are only two states, $k=31/128$ and $k=32/128$ playing an important role in the peak height whereas free fermion theory is correct in the regime where a quasi-continuum of wavevectors play a role in the singularity.  The way the peaks seen at finite size eventually merge into a smooth function with a singularity at $T_{c2}$ remains mysterious.  

\begin{figure}[h!]
    \centering
    \includegraphics[width=\columnwidth]{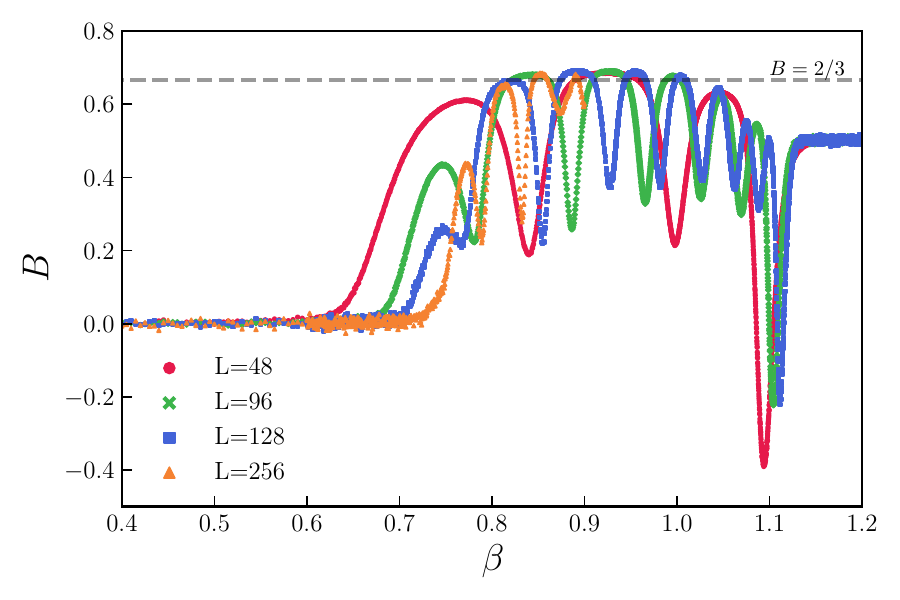}
    \caption{Binder cumulant, $B$, as a function of $\beta$ for sizes $L=48,96,128$, and $256$. The rise from zero is at larger $\beta$ for larger $L$.  
    Results for $L=256$ plotted up to $\beta=0.9$.
    }
    \label{fig:binder-cumulant}
\end{figure}

In an effort to identify the transition from the paramagnetic to the IC phase at $T_{c1}$ we measured the Binder cumulant of the overlap, $B$ (see Eq.~\eqref{eq:binder}). Figure \ref{fig:binder-cumulant} shows $B$ as a function of $\beta$ for several system sizes (the curve for $L=256$ ends at $\beta=0.9$).  As expected, $B$ is near zero in the high temperature phase and then the envelope of the curve rises to a value near $2/3$ in the finite-size IC phase but is punctuated by dips associated with the wavevector transitions. Our results are somewhat similar to those of Refs.~\cite{ matsubara_domain_2017} but with extra structure due to wavevector transitions and a much sharper dip at $T_{c2}$.  Note that the dips in $B$ correspond to the peaks in specific heat.  

In principle, $T_{c1}$ is marked by where $B$ rises above zero  but this rise strongly depends on system size, as is characteristic of a Kosterlitz--Thouless phase transition, so we cannot  make a quantitative estimate $T_{c1}$ from the data. Qualitatively, we see that as $L$ increases, small $\beta$ peaks disappear and the $\beta$ value where $B$ approaches a plateau value of about $2/3$  increases. Our results are consistent with the recent simulation result, $\beta_{c1}=0.86$ \cite{sato_matsubara_1999} and perhaps also with  $\beta_{c1}=0.96$ \cite{hu_resolving_2021}, though the latter requires that many of the small $\beta$ peaks disappear as $L$ increases.

\FloatBarrier

\subsection{Algorithm Performance}
\label{sec:algoresults}
In this section we compare the performance of the Metropolis, Wolff and TR algorithms. In the comparisons reported here, all algorithms have similar running times for a given system size.  Wolff and TR have the same number of sweeps per temperature step while Metropolis is given more sweeps to compensate for the greater simplicity of single spin versus cluster moves. In every case, a sweep is defined as changing or, in the case of Metropolis, attempting to change each spin in the population of replicas on average once. All results are computed from weighted averages of approximately 100 runs.  We show that for similar computational resources the TR algorithm performs best and also appears to have better scaling with system size than the other algorithms. We also report results on the variance of the free energy estimator that suggest that the TR algorithm is well-equilibrated for all the system sizes and temperatures studied except possibly $L = 128$ at the lowest temperatures.

\begin{figure}[h!]
    \centering
    \includegraphics[width=\columnwidth]{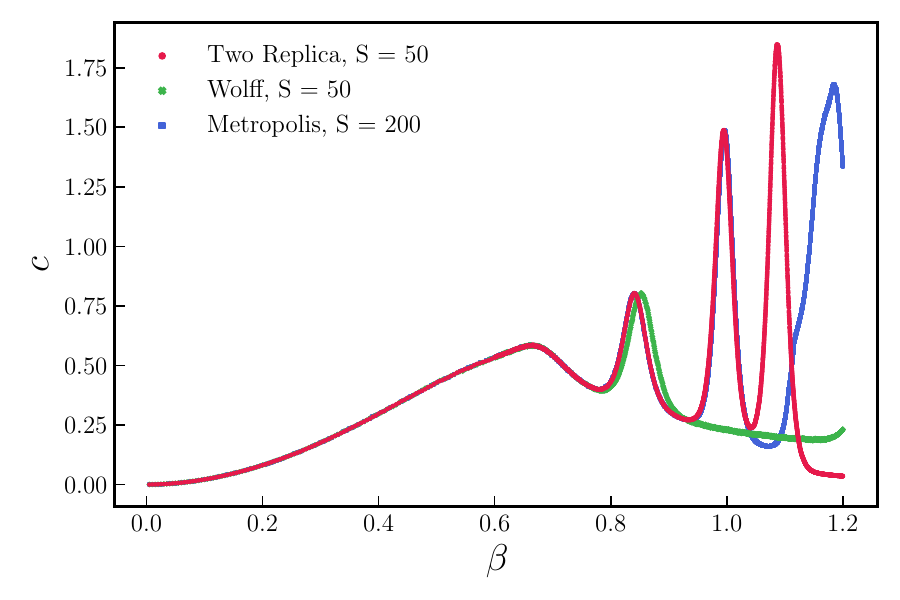}
    \hfill
    \includegraphics[width=\columnwidth]{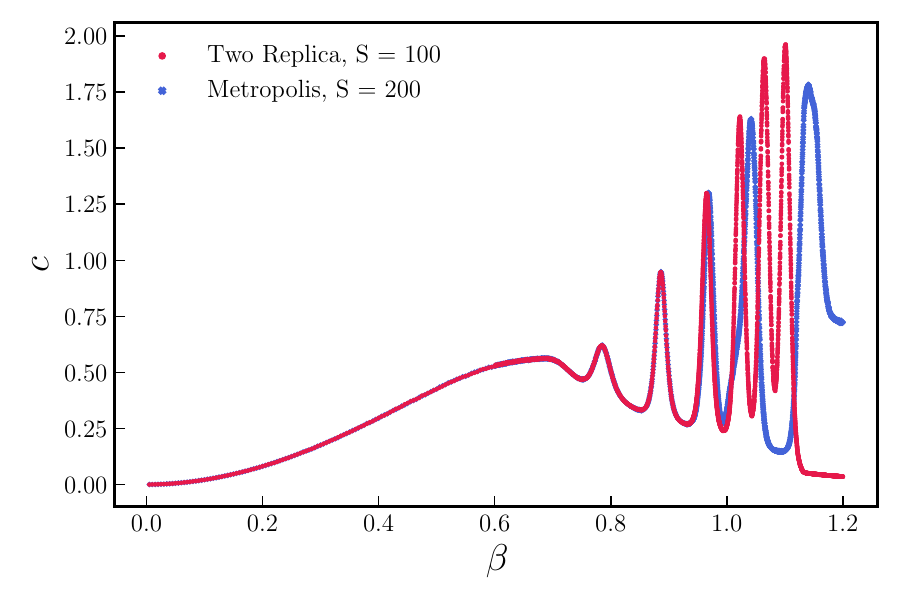}
    \caption{Specific heat curves comparing computational methods for $L = 48$ (top panel) and $96$ (bottom panel).  The red curve is the TR result and also the result of the other methods where they agree with TR. For $L = 48$ , Wolff  (green curve) deviates significantly from the TR results at $\beta=0.8$ and misses the final two specific heat peaks.  Metropolis (blue curve) finds all peaks but the last peak occurs at significantly higher $\beta$.  The number of sweeps for TR and Wolff are the same, while the number of sweeps for Metropolis is approximately four times larger. For $L = 96$, we compare only TR and Metropolis.  Metropolis misses the final specific heat peak and fails to find the $\Tt$ phase.}
    \label{fig:compareC48}
\end{figure}

Figure \ref{fig:compareC48} shows the specific heat as a function of inverse temperature for the three algorithms for sizes (a) $L=48$ and (b) 96. For size 48, the three algorithms give nearly the same results until $\beta=0.8$, where Wolff falls badly behind and then fails to reproduce the last two peaks or find the $\Tt$ phase.  Metropolis does better but is very late finding the last peak and the $\Tt$ phase.  For size 96, Metropolis fails to find the final peak or the $\Tt$ phase (the dominant wavevector $k=1/4$ is never observed).  These results show that Wolff performs poorly in the $\kappa> 1/2$ region of the phase diagram. Metropolis performs better than Wolff but is much less efficient than TR even for $L=48$. 

\begin{figure}[h!]
    \centering
    \includegraphics[width=\columnwidth]{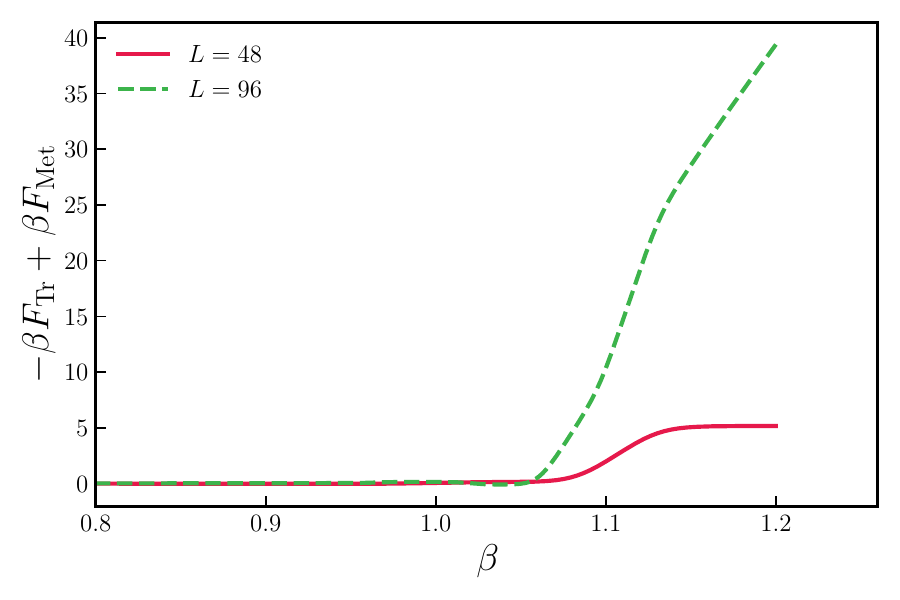}
    \caption{Free energy differences between the TR and Metropolis algorithms as a function of $\beta$, for $L = 48$ (solid red) and $96$ (dotted green).  Positive values indicate that the TR algorithm has a lower free energy.}
    \label{fig:compareF48}
\end{figure}

To lend weight to the idea that the Metropolis algorithm scales less well with $L$ than the TR algorithm we compare their free energy estimators.  
Figure \ref{fig:compareF48} shows the difference in the weighted average free energy estimator between TR and Metropolis as a function of $\beta$ with positive values indicating lower free energy for the TR algorithm.  This plot demonstrates that, indeed, the TR algorithms finds states of significantly lower free energy and that the difference increases with system size.  The very large discrepancy for $L=96$ beyond $\beta_{c2}\approx 1.1$ reflects the failure of the Metropolis algorithm to find the $\Tt$ phase.  

\begin{figure}[h!]
    \centering
    \includegraphics[width=\columnwidth]{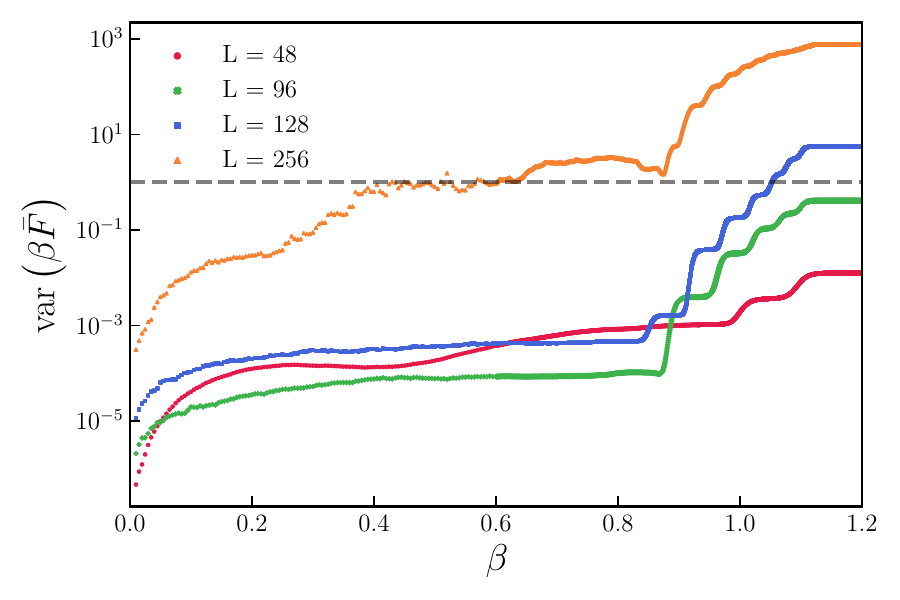}
    \caption{Variance of the weighted free energy estimator $\beta \bar{F}$, from bootstrapping, for the TR algorithm. The region below the dashed line at 1 is expected to be well-equilibrated.}
    \label{fig:varF}
\end{figure}

While a comparison of free energies shows that the TR algorithm is better equilibrated than the Metropolis algorithm, it does not show that the former has actually converged to equilibrium.  Figure \ref{fig:varF} shows the variance of $\bar{F}$ as a function of $\beta$ for all system sizes studied. A well-studied criterion for equilibration \cite{WaMaKa15b} is $\mathrm{var}(\beta  \bar{F})<1$. We see that all runs are well-equilibrated except $L=128$, which may not be fully equilibrated in the IC region, and $L=256$, which is poorly equilibrated for $\beta>0.9$.   It is possible that specific heat peak heights occur slightly late in $\beta$ for $L=128$.  In principle, the free energy variance may give a false signal of equilibration if important regions of phase space are missed by all runs in the weighted average. One promising sign that this problem does not occur here is that TR finds the $\Tt$ state for all runs for all system sizes, including $L=256$. 

The scaling with system size, $L$, of the number of TR sweeps, $MRS$, needed to reach equilibrium at low temperatures can be crudely estimated by fitting  values of $MRS\mathrm{var}(\beta  \bar{F})$ as a function of $L$ at the lowest temperature ($\beta =1.2$).  An exponential fit for sizes $L=48, 96, 128$ of the form $S_0 e^{L/L_0}$  yields $S_0=260$ and $L_0=13$.  Note that exponential scaling in $L$ is better than $L^2$, though the latter cannot be excluded given the limited range of $L$.  Furthermore, since even $L=256$ is much too small to see the expected asymptotic behavior of the specific heat, we cannot make claims about the asymptotic scaling of equilibration work.

\begin{figure}[h!]
    \centering
    \includegraphics[width=\columnwidth]{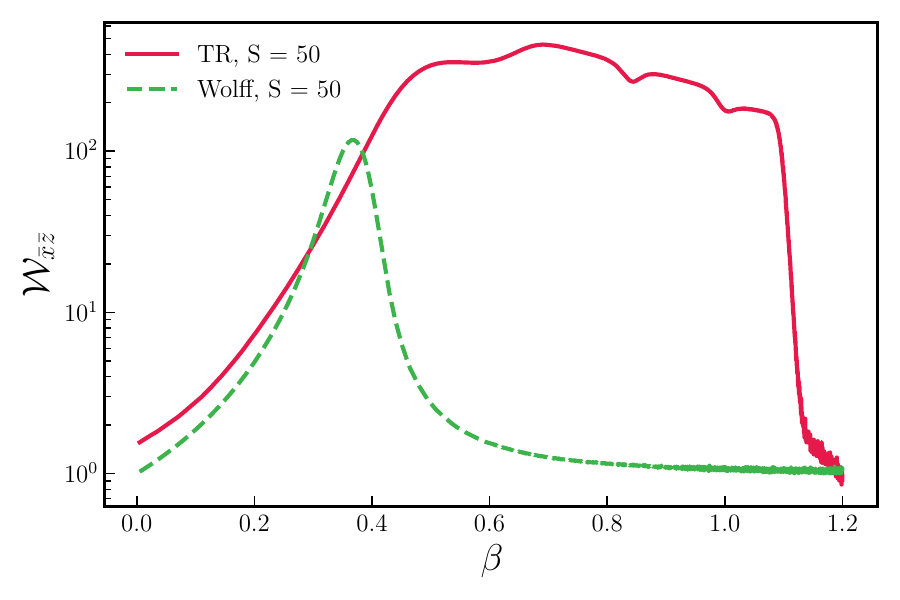}
    \caption{Average non-wrapping cluster size, $\Wnxnz$ as a function of $\beta$ for both the Wolff (green) and TR (blue) algorithms for size $L=48$. }
    \label{fig:no-wrap}
\end{figure}

Although both the Wolff and TR algorithms feature cluster moves, it is clear that TR clusters are far more efficient in achieving equilibration.  Some understanding of this difference is apparent from Fig.~\ref{fig:no-wrap}, which shows the average size, $\Wnxnz$, of non-wrapping  TR and Wolff clusters as a function of $\beta$ for size $L=48$. The non-wrapping Wolff clusters are large near an unphysical percolation transition deep in the disordered phase but are rare and almost always singletons in the region of interest encompassing the IC phase. For $\beta>0.6$ most Wolff cluster flips perform a global spin flip so the algorithm is quite inefficient compared to both Metropolis and TR. On the other hand, non-wrapping TR clusters are large throughout the IC phase, and then rapidly become small after the transition to the ordered phase.  Thus, the TR algorithm can make large changes to the system in the IC phase.  

\FloatBarrier

\subsection{Efficiency of Two-replica Cluster Moves}
\label{sec:TRflips}
We have seen that the two-replica cluster algorithm (TR) performs better than the single spin-flip (Metropolis) or single-replica cluster  (Wolff) algorithms when all algorithms are combined with population annealing.  In this section we provide an explanation for the much greater efficiency of the combination of two-replica and Metropolis moves together with population annealing. The key observation is that a two-replica cluster flip in the IC region is capable of moving defect lines between replicas. Although the total number of defect lines in the population is conserved by two-replica cluster flips, population annealing can change the number of defect lines in the population via resampling. 

\begin{figure}[h!]
    \centering
    \includegraphics[width=1.0\linewidth]{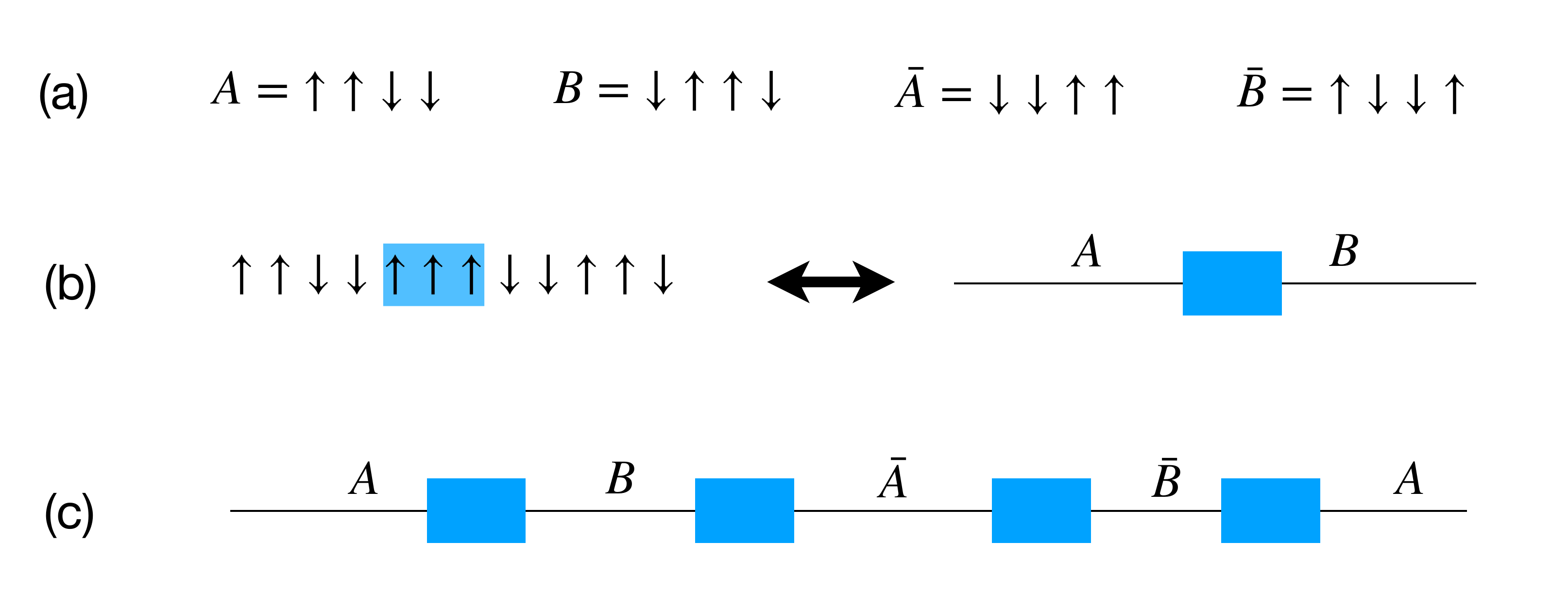}
    \caption{ (a) Labels for the four $\Tt$ phases.   (b) $A$ and $B$ regions separated by a three-spin defect and, on the right, a simplified representation of the $A$ and $B$ regions (of arbitrary length) with the three-spin defect represented as a blue rectangle. (c) A sequence of three-spin defects and the required types of $\Tt$ phases between them. Note that the three-spin defects may be composed of either up or down spins. }
    \label{fig:cluster1}
\end{figure}

To simplify the analysis, we first consider one-dimensional ANNNI chains at low temperature in configurations with $\Tt$ regions separated by widely spaced  three-spin defects.  These are the domain states discussed in Ref.~\cite{matsubara_domain_2017} and we use a similar notation to discuss them.  In Fig.~\ref{fig:cluster1}a we label the four $\Tt$ states as $A$, $B$, $\bar{A}$, and $\bar{B}$.  Figure \ref{fig:cluster1}b illustrates how a three-spin defect changes the $\Tt$ phase.  On the right side of Fig.~\ref{fig:cluster1}b we introduce a simplified representation of two distinct $\Tt$ regions separated by a three-spin defect. Figure \ref{fig:cluster1}c shows a configuration starting and ending in the type $A$ $\Tt$ phase with four three-spin defects separating regions (of arbitrary length) of the other three phases.  Note that $\Tt$ regions separated by three-spin defects must always appear in the sequence $A \rightarrow B \rightarrow \bar{A} \rightarrow \bar{B}\rightarrow A$. Defects with either three up spins or three down spins induce the same change in phase and, more generally, a defect with $n$ identical spins will permute the phase through the $AB\bar{A}\bar{B}$ sequence by a cycle of length $(n-2) \mod{4}$.  This rule also holds for $n=1$.

\begin{figure}[h!]
    \centering
    \includegraphics[width=1.0\linewidth]{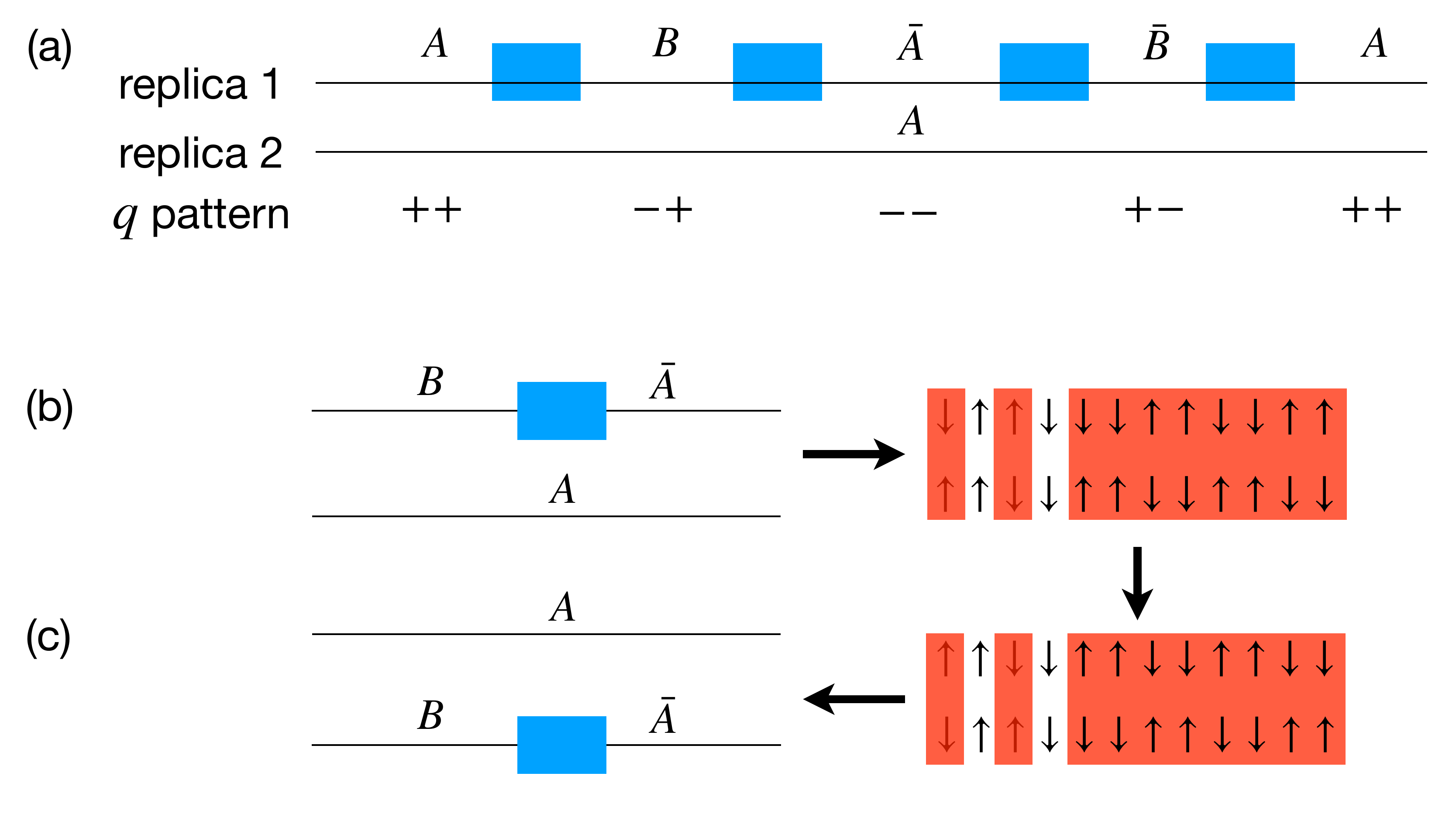}
    \caption{A pair of replicas and a cluster flip: (a) The initial state of the replicas with the overlap pattern in each region shown below. (b) Detail of the $B$-$\bar{A}$ interface with the low temperature, $q=-1$ cluster shown with red background. (c) Spin configuration after the cluster flip, note that the cluster flip interchanges the configurations of replicas 1 and 2, moving the the defect from 1 to 2.  }
    \label{fig:cluster2}
\end{figure}

Figure \ref{fig:cluster2}a shows a pair of replicas with replica 1 having four defects in the region of interest and replica 2 in the $A$ phase throughout the region of interest.  Beneath the schematic of the two replicas is the overlap pattern in the various regions:  the overlap alternates in regions where the letter type between replicas differs  (e.g. $\bar{A}B$) and the overlap is constant in regions where the letter type is the same (e.g.\ $\bar{A}A$).  

Now consider a low temperature two-replica cluster whose seed is at a $q=-1$ site. In a low temperature cluster, all satisfied bonds are occupied.  Figure \ref{fig:cluster2}b shows the region in replica 1, with phases $B$ and $\bar{A}$ separated by a three-spin defect.  A detail of the defect and the region immediately surrounding it is shown on the right with the cluster highlighted in red.  It is straightforward to check that at low temperatures the cluster will occupy all sites with $q=-1$ and is able to jump over three-spin defects.  Figure \ref{fig:cluster2}c shows the state of the system after the cluster flip with the result that replica 1 and 2 simply exchange their states in the region shown.  By the same reasoning, the full cluster flip interchanges the entire region shown in Fig.~\ref{fig:cluster2}a, which includes all four defects. Note that the cluster is bounded by the outer $AB$ and $\bar{B}\bar{A}$ three-spin defects since outside of these defects the overlap is positive. The end result is four defects have been moved from replica 1 to replica 2.

For the 2D model, three-spin defects become three-spin defect lines and the $\Tt$ regions become strips but the same ideas apply: defect lines and regions between them are interchanged between replicas by two-replica cluster flips.    

We have seen that two-replica cluster flips are able to effect global changes by interchanging large regions between replicas, bounded by defect lines. On the other hand, two-replica cluster flips by themselves cannot equilibrate the whole population because the number of defect lines is conserved by two-replica cluster flips. This weakness is remedied by embedding the two-replica algorithm in population annealing. If there are variations in the number of defect lines in the population, then PA will preferentially eliminate replicas with more defect lines so that defect lines can be efficiently annealed from the population as the temperature decreases.

\begin{figure}[h!]
    \centering
    \includegraphics[width=\columnwidth]{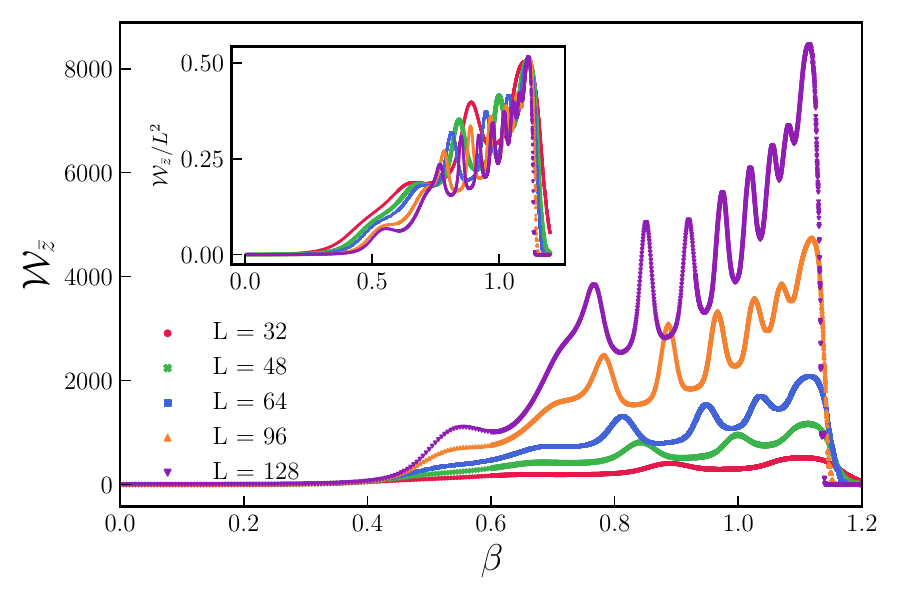}
    \caption{Average size of clusters that do not wrap in the $z$ direction, $\Wnz$, as a function of $\beta$ for different lattice sizes. Inset shows the same quantity divided by total number of spins, $L^2$.}
    \label{fig:nonzwrap}
\end{figure}

\begin{figure}[h!]
    \centering
    \includegraphics[width=\columnwidth]{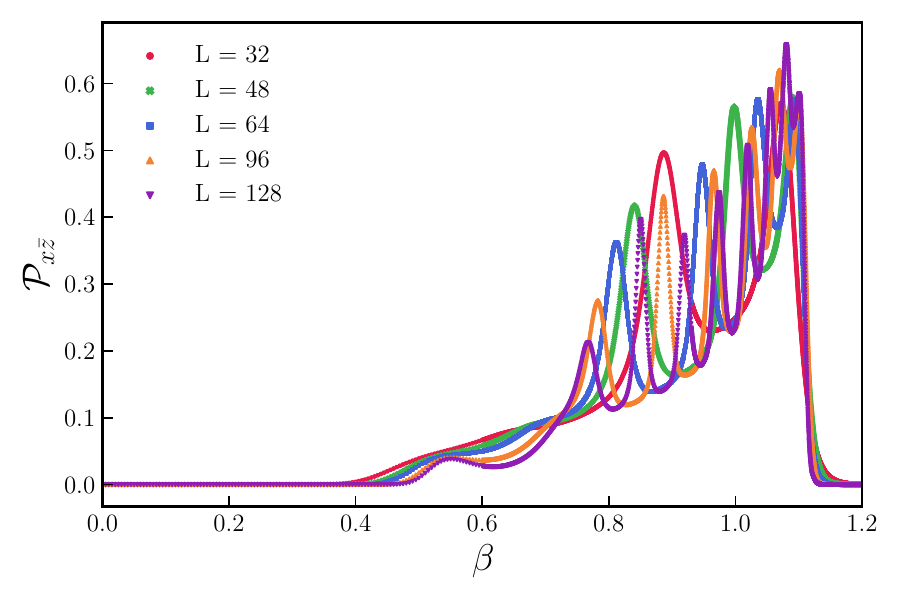}
    \includegraphics[width=\columnwidth]{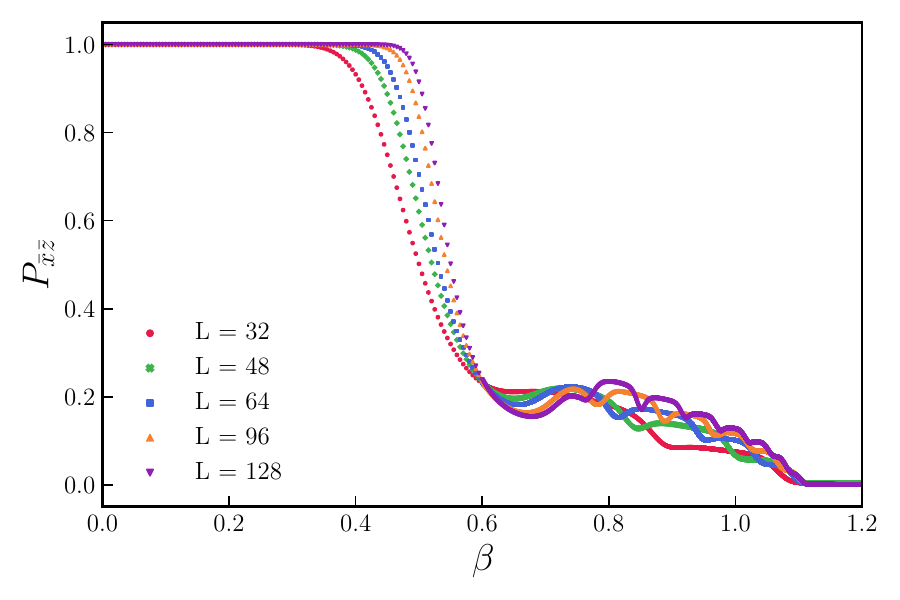}
    \caption{Probability of wrapping in only the $x$-axis, $\Pxnz$ (top panel), and probability of not wrapping, $\Pnxnz$ (bottom panel), for different system sizes.}
    \label{fig:finitecluster}
\end{figure}

To support the above theoretical picture, we studied the probabilities and average size of two-replica clusters with different wrapping conditions. We have already seen in Fig.~\ref{fig:no-wrap} that the average size of non-wrapping (finite) clusters is large until $T_{c2}$ when it falls sharply to one. Figure \ref{fig:nonzwrap} shows the average size, $\Wnz$, of clusters that do not wrap in the $z$ direction. We see that $\Wnz$ increases as $\beta$ increases before abruptly falling to one in the $\Tt$ phase. The inset of Fig.~\ref{fig:nonzwrap} shows for $\beta\gtrsim 0.75$ that $\Wnz$ scales roughly as the system size and reaches roughly $1/2$ the system size at the last peak before the $\Tt$ phase. Clusters that contribute to $\Wnz$ may or may not wrap in the $x$ direction. In order to disentangle these contributions, the probabilities of not wrapping in the $x$ direction and wrapping in the $x$ direction, while not wrapping in the $z$ direction, $\Pnxnz$ and $\Pxnz$, respectively, are shown in Fig.~\ref{fig:finitecluster}, lower and upper panels, respectively. We see that near $T_{c2}\approx 1.1$ almost all clusters wrap and that roughly half of these clusters wrap only in the $x$ direction. Clusters that wrap only in the $x$ direction are of the type discussed above that are confined by defect lines and are capable of moving groups of four defect lines between replicas. The fact that these clusters are common in the IC phase suggests that the mechanism discussed above is effective. On the other hand, near and above the transition, fully non-wrapping clusters are quite rare ($\Pnxnz \approx 0$), which explains why it is also important to include single-spin-flip dynamics in the full TR algorithm.

Figure \ref{fig:finitecluster}, lower panel, shows a crossing in $\Pnxnz$ for different system sizes that suggests a percolation transition at $\beta \approx 0.6$. We do not believe that this transition corresponds to a thermodynamic transition and speculate that it corresponds to the appearance (in the all-cluster picture) of two wrapping clusters of nearly equal size and opposite overlap, as seen in two-replica simulations of spin glasses \cite{MaStNe07}.

\section{Discussion}
\label{sec:discuss}

We have introduced a new computational method for simulating the ANNNI model and used it to study the 2D model with periodic boundary conditions for $\kappa=0.6$, where the low temperature phase is layered. The finite-size incommensurate floating phase is characterized by a sequence of well-resolved, sharp specific heat peaks culminating in a transition to the low temperature $\Tt$ phase. At the largest sizes studied, we see  hints of a cross-over to a smooth specific heat function culminating in a singularity at the transition but the nature of this crossover remains unclear.  The transition temperature, $T_{c2}=0.91$, is consistent with previous results.  Other features of the finite-size IC phase contrast with those obtained from infinite aspect ratio transfer matrix methods and finite aspect ratio simulations in free boundary conditions, suggesting the need for additional studies of finite-size behavior as a function of boundary conditions and aspect ratios.  

The TR algorithm introduced here combines two-replica cluster flips, single spin flips and population annealing and was shown to be very effective at equilibrating the 2D ANNNI model.  We have shown that the new algorithm is much more efficient than either the Metropolis algorithm or the Wolff algorithm when these are combined with population annealing.  It would be interesting to see how the two-replica method compares with the cluster heat bath method \cite{matsubara_sato_koseki_1997}. We have argued that the effectiveness of the algorithm in the finite-size incommensurate floating results from the ability of the two-replica cluster flips to transfer groups of four defect lines between replicas. This suggests that the same algorithm may be useful for the 3D ANNNI model and related spin models that feature layered structures due to competing short-range and long-range interactions.  

\acknowledgements
We thank Patrick Charbonneau and Mingyuan Zheng for useful discussions.  The simulations were carried out on the Unity Cluster of the Massachusetts Green High Performance Computing Center.
\bibliography{references}

\end{document}